\documentclass{webofc}

\usepackage[varg]{txfonts}   
\usepackage{hyperref}
\usepackage{url}
\hypersetup{colorlinks=true,citecolor=blue,urlcolor=blue,linkcolor=blue}
\begin{document}
\title{GlideinBenchmark: collecting resource information to optimize
provisioning}

\author{
        \firstname{Marco} \lastname{Mambelli}\inst{1}\fnsep\thanks{\email{marcom@fnal.gov}} \and
        \firstname{Shrijan} \lastname{Swaminathan}\inst{2}
}

\institute{
    Fermi National Accelerator Laboratory,
    PO Box 500, Batavia IL 60510-5011
\and
    Purdue University
          }

\abstract{
Choosing the right resource can speed up job completion, better utilize the available hardware, and visibly reduce costs, especially when renting computers in the cloud. This was demonstrated in earlier studies on HEPCloud. However, the benchmarking of the resources proved to be a laborious and time-consuming process. This paper presents GlideinBenchmark, a new Web application leveraging
the pilot infrastructure of GlideinWMS to benchmark resources, and it shows how to use the data collected and published by GlideinBenchmark to automate the optimal selection of resources.
An experiment can select the benchmark or the set of benchmarks that most closely evaluate the performance of its workflows. GlideinBenchmark, with the help of the GlideinWMS Factory, controls the benchmark execution. Finally, a scheduler like HEPCloud’s Decision Engine can use the results to optimize resource provisioning.
}
\maketitle
\section{Introduction}
\label{intro}
GlideinBenchmark works within a pilot-based workload management system to submit benchmarks and collect results that can be used to improve resource provisioning.

The next section will introduce pilot-based workload management systems like GlideinWMS and HEPCloud. The following will present how benchmarks have been used to improve manual or automated provisioning. 
The following two sections will describe the GlideinBenchmark architecture and how it interacts with a pilot-based system. 
Finally, this paper will conclude with use considerations and ideas for the future evolution of GlideinBenchmark.

\section{Pilot-based workload management systems}
\label{pilot-wms}
\begin{figure}[htpb]
\centering
\includegraphics[width=8cm,clip]{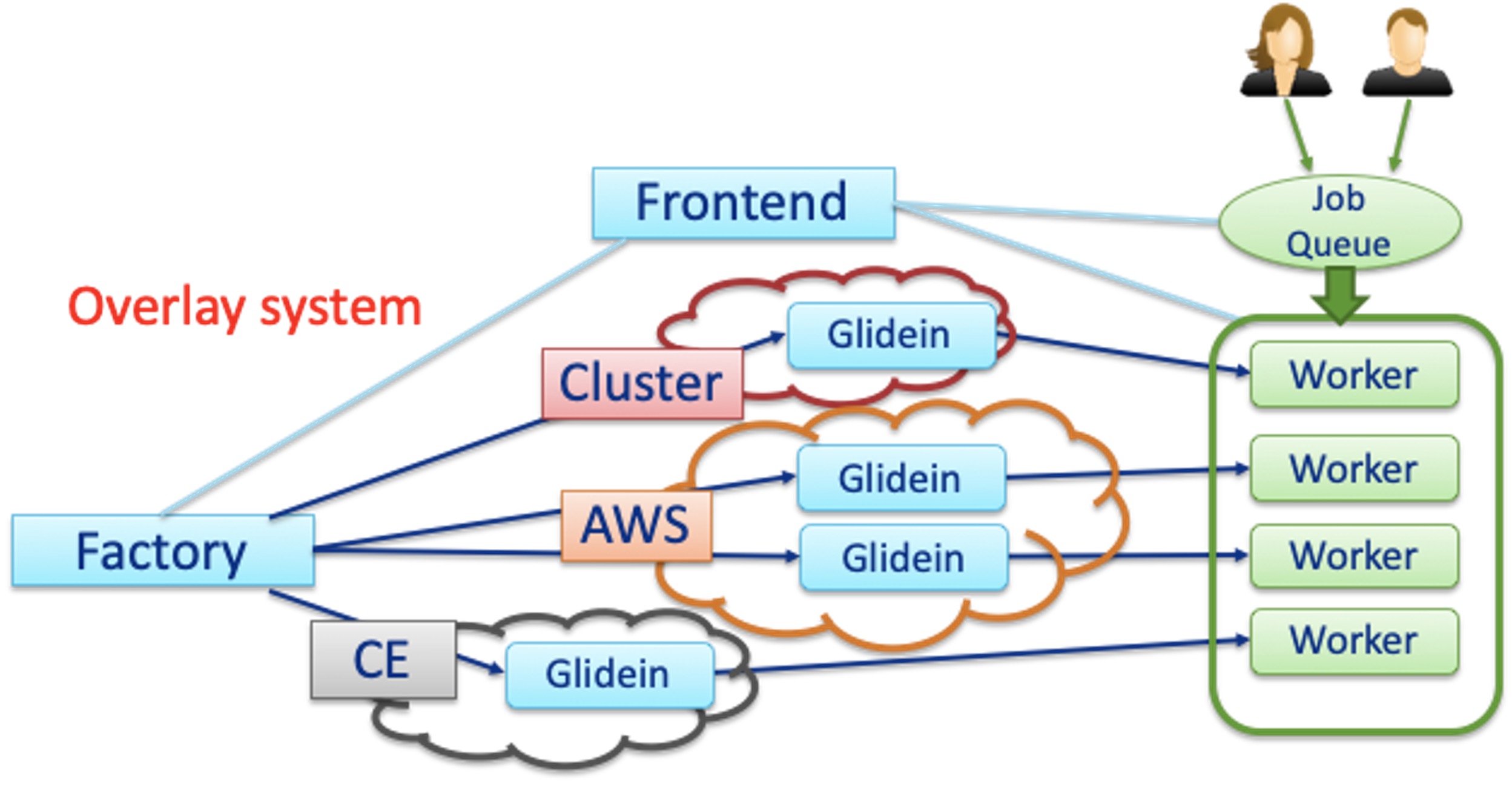}
\caption{GlideinWMS system. GlideinWMS components are in blue, the User Pool is in Green, and the computing resources are in other colors.}
\label{fig-gwms}       
\end{figure}

GlideinWMS~\cite{glideinwms,gwms-sw} is a pilot and pressure-based Workload Management System (WMS) provisioning computing resources in a distributed environment. HEPCloud\cite{hepcloud} is also a pilot-based WMS, but thanks to its Decision Engine\cite{hepcloud-de-sw}, it can use more complex resource-provisioning strategies. Their users can request one or more customized elastic HTCondor Software Suite (HTCSS)\cite{htcondor} clusters, User Pools, in green in figure~\ref{fig-gwms}, where the users run their computations.
To provision the elastic cluster, GlideinWMS sends to a variety of computing resources shown in figure~\ref{fig-gwms} Glideins, also called pilot jobs to distinguish them from the scientific computations, the user jobs.
GlideinWMS has been used for more than 10 years and is used at scale in production by many collaborations, including the Compact Muon Solenoid (CMS) experiment, many Fermilab experiments, and the OSG. 
Most scientists will not use directly GlideinWMS or the clusters it provides, they will interact instead with the various tools or portals like CRAB, JobSub, or OSG-Connect, provided by the scientific collaborations.

The key component is the Glidein, the pilot job. It is a program sent to many resources that match the preliminary requirements of the user jobs. It tests and sets up each computing resource to run the user jobs. 
It can auto-detect and report node resources like CPU cores, memory, disk, and GPUs, can install common resources like a container runtime or a distributed file system, and provides monitoring and audit information.
It finally joins the User Pool to run one or more user jobs, in parallel and in sequence, depending on the needs and availability.
It is common for the Glidein to run user jobs inside containers. Container images, the templates used to start containers, are downloaded or read from the CernVM-FS~\cite{cvmfs}, a distributed file system available on most resources.

The Factory and clients like the Frontend or HEPCloud's Decision Engine complete the GlideinWMS system. For clarity, in this paper, we'll consider a system with one Frontend, one Factory, and their Glideins. Actual deployments may include multiple clients, differing in how they calculate the requests for the Factory, and multiple Factories, providing a redundant distributed system.

The Factory is in charge of submitting Glideins to the different Compute Entrypoints, CEs. 
It knows how to reach each computing resource, which collaborations are supposedly supported, which protocols and authentication methods are supported, and if there are throttling requirements. It submits Glideins to the Compute Entrypoints, maintaining on each one the pressure, i.e., the number of queued and running Glideins, requested by its clients.
The Factory monitors the Glideins and hosts a secure mailbox to exchange requests and status messages with the clients.

The Frontend and other clients are aware of the users' requests and the running and queued Glideins that can be used for those requests, they receive resource status information from the Factory, and they use heuristics to update the requests to the Factories so that all the user jobs can run promptly, all limits and policies are respected, and no resources are wasted.
The Frontend is generally operated by the scientific experiments or on their behalf and implements their policies for resource provisioning and job priorities.

\section{Benchmark-based resource selection}
\label{benchmark}
\begin{figure}[h]
\centering
\includegraphics[width=6cm,clip]{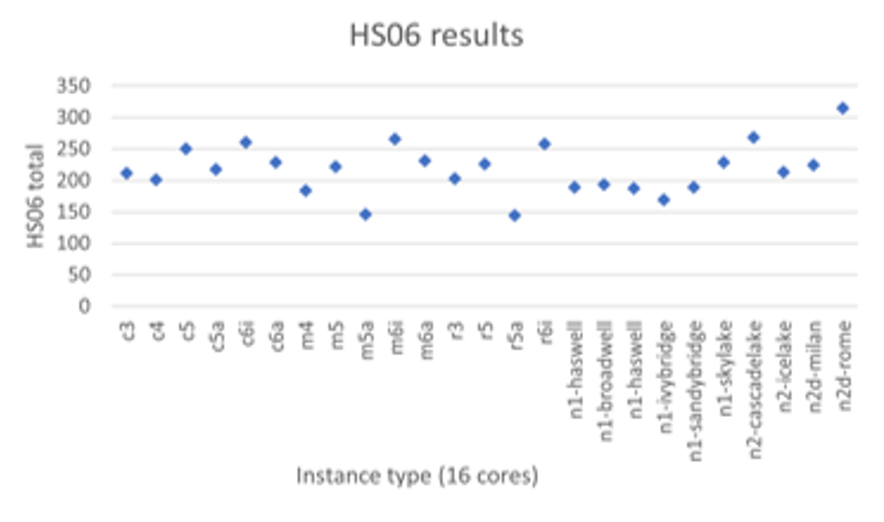}
\includegraphics[width=6cm,clip]{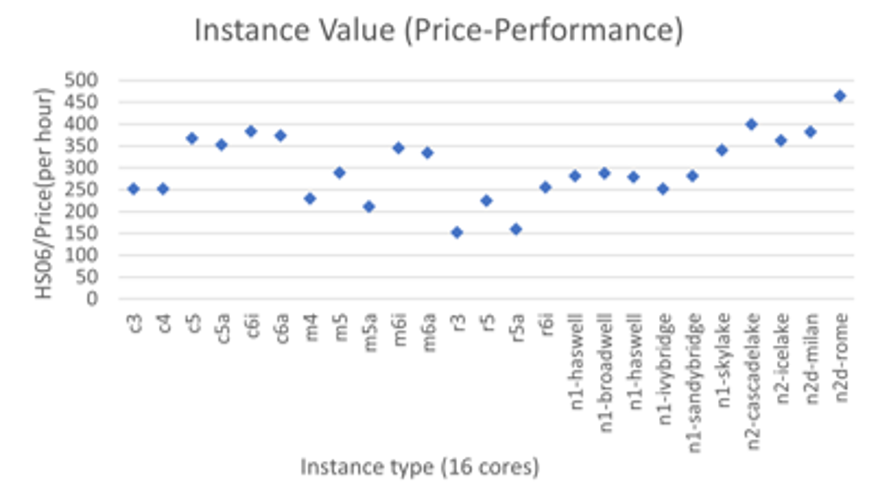}
\caption{Benchmark results - Higher generation CPUs perform better, even more so when the price-performance ratio is considered}
\label{fig-bench}       
\end{figure}

HEP software performance can vary greatly on different architectures. 
HEPCloud Decision Engine is a client similar to the Frontend, speaks the same protocol to request Glideins from a Factory, but has more complex algorithms to select the resources. It has been used to minimize the cost-over-performance ratio when running on Amazon's cloud (EC2) \cite{hepc-aws-bidding}. The code used a thorough manual benchmarking of the different EC2 instances to select the best mix.
Manual benchmarks like the ones in figure~\ref{fig-bench} are very useful for provisioning resources, but are time-consuming to perform and become quickly obsolete when new processors or architectures are available.
Automating the benchmark collection would maintain the benefit of fresh data and remove the labor-intensive process. GlideinBenchmark was born to answer this need.

\section{GlideinBenchmark}
\label{gbench}
\begin{figure}[h]
\centering
\includegraphics[width=12cm,clip]{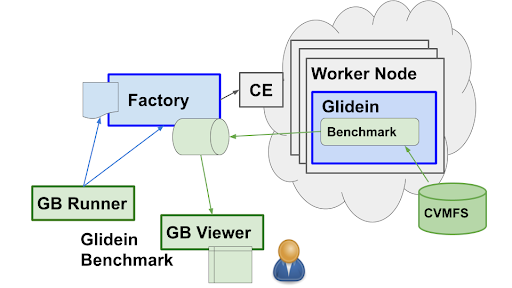}
\caption{GlideinBenchmark architecture. The Runner controls the benchmark execution via the Factory. The viewer stores and displays the benchmark results received from the Glideins.}
\label{fig-arch}       
\end{figure}

Figure~\ref{fig-arch} shows in green the two independent components of GlideinBenchmark. Two web dashboards in Python, the Runner and the Viewer.  Both have been designed to be as lean as possible while giving options for added features.  They can run on the GlideinWMS Factory host or separately.
GlideinBenchmark relies on the GlideinWMS Factory for the actual execution of the benchmarks.
Benchmarks are run on Apptainer~\cite{apptainer} containers, and their images are distributed via the CernVM-FS.

Python was chosen because it is the main language in GlideinWMS, all components use it, and it is well known by GlideinWMS developers.
The Factory is a well-established system for running jobs on the managed resources. It would be a duplication of effort to write a new system to run the benchmarks.
Running the benchmarks in containers eases the configuration and broadens the availability. The containerization software Apptainer is widely used by GlideinWMS stakeholders and has an established track record running without admin privileges, which is necessary because Glideins run unprivileged on the resources. 

The design of this project allows the Web server (Viewer) and the control system (Runner) to be completely isolated, assuming they can interact with the same Factory and its HTCondor system.

\subsection{GlideinBenchmark Control System (Runner)}
\label{gb-runner}
The Runner controls the behavior of the Factory. Specifically, it can change the Factory configuration to enable benchmarks, trigger the use of the new configuration, and trigger the submission of special Glideins running ad-hoc benchmarks. To do so, it needs to have write access to the Factory configuration files, to issue system commands like the Factory reconfiguration, and to send queries and commands to the Factory HTCondor system to submit the special benchmark-Glideins.

\subsection{GlideinBenchmark Web Server or View System (Viewer)}
\label{gb-viewer}
The Viewer shows the status of the running benchmarks, if any, and the results collected by the past benchmarks. To do so, it needs to read data stored in the Factory and query its HTCondor system.

\subsection{Benchmark execution}
\label{gb-exe}
Benchmarks are executed by the Glideins using Apptainer containers running on the resources.

To run benchmarks, all resources must have Apptainer and the CernVM-FS, but this is already a requirement to run almost all user jobs using GlideinWMS.
In actuality, both of these requirements can be relaxed. The most recent versions of the HTCondor tarball included in the Glideins also contain Apptainer, so it does not need to be installed on the worker nodes. And some worker nodes provide in their local storage replicas of the container images distributed by OSG, like the benchmark images, so they can be accessed without CernVM-FS.

\section{Integration with provisioning}
\label{integration}
The Runner is controlled by users via the Web UI or CLI. It can update the Factory configuration files, issue a reconfig or update commands (i.e., trigger the use of the new configuration), and launch Glideins customized to run the desired benchmarks using the Factory API to submit individual Glideins. There is no direct interaction with the Viewer or other components.
The Viewer gets its information about the number of benchmarks running or queued from HTCondor and the other information, like the results of past tests, from Factory-collected log files like the Glidein logs. This information can be preprocessed to ease its consumption and speed up the retrieval. 

A simple deployment could see GlideinBenchmark running on the same server as the GlideinWMS Factory, using the same Unix user, to easily update the configuration file, control the Factory, and issue HTCondor queries and commands.

The user interactions are via a Web GUI or using a REST API.
The Factory interactions use the GlideinWMS XML configuration and API
Benchmark results are transferred leveraging the GlideinWMS Glideins standard error, a mechanism used already to transfer other files.

HEPCloud's Decision Engine and other Factory clients can access the benchmark results via the same REST API used by operators and can update their evaluation of the different resources. They would always have fresh information to guide the resource provisioning.

\section{Conclusions}
\label{conclusion}
GlideinBenchmark is a useful addition that automates the collection of benchmarks that resource provisioners like HEPCloud's Decision Engine can use.
It leverages components already present in the GlideinWMS framework, like the Glideins, Apptainer, and the CernVM-FS.

GlideinWMS Factories can have a few hundred entries. GlideinBenchmark will need to be tested at scale, using on the order of a thousand different Factory entries and be able to complete a test of the whole system within a few hours. We do not expect the benchmark to run more than once daily on a given resource. It could also be triggered only for new resources, and an intelligent sampling could avoid running the benchmark on all resources. It is important to guarantee minimum overhead and no interference with the jobs' execution. Future work will have to select the best benchmarks to evaluate the resources.

\section{Acknowledgments}
\label{ack}
The authors’ work was performed using the resources of the Fermi National Accelerator Laboratory (Fermilab), a U.S. Department of Energy, Office of Science, HEP User Facility. Fermilab is managed by Fermi Forward Discovery Group, LLC, acting under Contract No. 89243024CSC000002. 

\bibliography{gwms-hepcloud} 

\begin{thebibliography}{8}

\bibitem{glideinwms}
I.~Sfiligoi, glideinwms—a generic pilot-based workload management system, Journal of Physics: Conference Series \textbf{119}, 062044 (2008). \doiwoc{10.1088/1742-6596/119/6/062044}

\bibitem{gwms-sw}
I.~Sfiligoi, M.~Mambelli, P.~Mhashilkar, D.~Box, M.~Mascheroni, K.~Larson, B.~Holzman, J.~Weigand, A.~Tiradani, H.W. Kim et~al., glideinwms/glideinwms: Glideinwms 3.10.5 (2023), \urlstyle{tt}\url{https://doi.org/10.5281/zenodo.8383959}

\bibitem{hepcloud}
{Mhashilkar, Parag}, {Altunay, Mine}, {Berman, Eileen}, {Dagenhart, David}, {Fuess, Stuart}, {Holzman, Burt}, {Kowalkowski, James}, {Litvintsev, Dmitry}, {Lu, Qiming}, {Moibenko, Alexander} et~al., Hepcloud, an elastic hybrid hep facility using an intelligent decision support system, EPJ Web Conf. \textbf{214}, 03060 (2019). \doiwoc{10.1051/epjconf/201921403060}

\bibitem{hepcloud-de-sw}
P.~Riehecky, K.~Knoepfel, D.~Litvintsev, P.~Mhashilkar, M.~Mambelli, V.D. Benedetto, P.~Gartung, S.~Bhat, L.~Goodenough, B.~Coimbra et~al., {HEPCloud/decisionengine: HEPCloud decisionengine 2.0.2} (2022), \urlstyle{tt}\url{https://doi.org/10.5281/zenodo.7108649}

\bibitem{htcondor}
T.~Tannenbaum, D.~Wright, K.~Miller, M.~Livny, in \emph{Beowulf Cluster Computing with {L}inux}, edited by T.~Sterling (MIT Press, 2001)

\bibitem{cvmfs}
{CernVM File System}, last Accessed: February 28, 2025, \urlstyle{tt}\url{https://cernvm.cern.ch/fs/}

\bibitem{hepc-aws-bidding}
H.~Wu, S.~Ren, S.~Timm, G.~Garzoglio, S.Y. Noh, Experimental Study of Bidding Strategies for Scientific Workflows using AWS Spot Instances (2015)

\bibitem{apptainer}
Apptainer - portable, reproducible containers, last Accessed: February 28, 2025, \urlstyle{tt}\url{https://apptainer.org/}

\end{thebibliography}
\end{document}